\documentstyle[sprocl,epsfig,amssymb]{article}

\def\be{\begin{equation}}
\def\ee{\end{equation}}
\def\bea{\begin{eqnarray}}
\def\eea{\end{eqnarray}}
\newcommand{\bx}{{\bf x}}

\newcommand{\bp}{{\bf p}}

\newcommand{\beq}{\begin{equation}}
\newcommand{\eeq}{\end{equation}}

\newcommand{\Tr}{\hbox{Tr}}
\newcommand{\C}{{\cal C}}
\newcommand{\rmd}{{\rm d}}

\newcommand{\bean}{\begin{eqnarray*}}
\newcommand{\eean}{\end{eqnarray*}}

\begin{document}

\title{QUANTUM FIELDS FAR FROM EQUILIBRIUM AND THERMALIZATION}

\author{
J. BERGES}

\address{Institute for Theoretical Physics, University of Heidelberg,\\
Philosophenweg 16,
69120 Heidelberg, Germany\\E-mail: j.berges@thphys.uni-heidelberg.de}


\maketitle\abstracts{I review the use of the
$2PI$ effective action in nonequilibrium quantum field \mbox{theory}.
The approach enables one to find approximation schemes which circumvent
long-standing problems of non-thermal
or secular (unbounded) late-time evolutions encountered
in standard loop or $1/N$ expansions of the $1PI$ effective action.  
It is shown that late-time thermalization can be described 
from a numerical solution of the three-loop $2PI$ effective action 
for a scalar $\phi^4$--theory in $1\!+\!1$ dimensions
(with J{\"u}rgen Cox, hep-ph/0006160).  
Quantitative results far from equilibrium beyond the weak coupling
expansion can be obtained from the $1/N$ expansion of the $2PI$ 
effective action at next-to-leading order (NLO),  
calculated for a scalar $O(N)$ symmetric quantum field theory
(hep-ph/0105311). Extending recent calculations in classical field theory 
by Aarts et al.\ (hep-ph/0007357) and by Blagoev et al.\ (hep-ph/0106195) 
to $N>1$ we show that the NLO 
approximation converges to exact (MC) results 
already for moderate values \mbox{of $N$} 
(with Gert Aarts, \mbox{hep-ph/0107129}).
I comment on characteristic time scales in scalar quantum field
theory and the applicability of classical field theory for sufficiently
high initial occupation numbers.
}

\section{Introduction}

Nonequilibrium dynamics of quantum fields is much less well understood 
than its thermal equilibrium limit. Important progress has been 
achieved for systems close to equilibrium, including 
(non)linear response techniques, gradient expansions or effective 
descriptions based on a separation of scales in the
weak coupling limit \cite{Bodeker:2001pa}. Current and upcoming 
heavy ion collision experiments provide an important motivation to find 
controlled approximation schemes which yield a quantitative description 
of {\bf far-from-equilibrium} phenomena. Other applications 
include (p)reheating at the end of inflation in the early 
universe, the physics of  baryogenesis or measurements of
the dynamics of Bose-Einstein condensation. 

To study nonequilibrium phenomena
no other dynamics than the known Hamiltonian
time-evolution of quantum fields is necessary. In particular,
the observed macroscopic, effectively dissipative behavior or 
thermalization has to arise from the underlying reversible quantum dynamics.
Addressing these questions one has to find 
approximation schemes which respect all symmetries of the underlying 
theory, such as time-reflection symmetry. Far from equilibrium
the dynamics can involve very different length scales and a priori
there is typically no effective description based on a clear 
separation of scales which is valid at all times. Similarly, 
the relevant renormalized interactions between modes can become
large at intermediate times and
expansions based on weak coupling arguments are limited.

Here we report on recent progress to find suitable approximation 
schemes for far-from-equilibrium dynamics  
based on the two-particle irreducible ($2PI$) generating functional 
for Green's functions \cite{Cornwall:1974vz}.
The approach does not employ the loop expansion of the $2PI$ effective 
action relevant at weak couplings \cite{Cornwall:1974vz,Calzetta:1988cq},
and extends first calculations of late-time thermalization
for quantum fields solving the three-loop 
$2PI$ approximation \cite{Berges:2000ur,Aarts:2001qa}.
To obtain a small nonperturbative expansion parameter for
far-from-equilibrium dynamics we consider a systematic 
{\bf 1/N expansion of the 2PI effective action}  
\cite{Berges:2001fi}. Since the truncation error may be controlled 
by higher powers of $1/N$ this expansion is not restricted to weak 
couplings or situations close 
to equilibrium. The next-to-leading order (NLO) approximation
has been solved numerically for a scalar $O(N)$ symmetric quantum 
field theory in $1\!+\!1$ dimensions \cite{Berges:2001fi,AB2}.
The $1/N$ expansion of the $2PI$ effective action in the NLO
approximation is equivalent to the ``BVA'' approximation\footnote{The name 
``Bare-Vertex-Approximation'' refers to
an ansatz on the level of a specific 
Schwinger-Dyson equation \cite{Mihaila:2001sr}. Note that the four-vertex 
gets renormalized at next-to-leading order $2PI$ \cite{Berges:2001fi}.} 
discussed earlier for quantum oscillators in Ref.\ \cite{Mihaila:2001sr}.  
Recently, this approximation has been studied for a one-component
classical field theory in Ref.\ \cite{Blagoev:2001ze} and compared
with exact (MC) results by numerically integrating the classical
evolution equations and sampling over initial conditions.\cite{Test}
$\!$ Extending
this analysis to the $N$-component classical $O(N)$ model 
one can show \cite{AB2} that the $1/N$ expansion at NLO yields systematically,
by increasing $N$, quantitatively precise 
results already for moderate values of $N$. 

Controlled computational methods 
for the approximative solution of non\-equilibrium dynamics
are limited so far. Leading order large-$N$ or mean field type approximations
neglect scattering and are known to fail to describe 
thermalization \cite{LOinh}.
Standard (finite-loop) perturbative descriptions \cite{Andreas}
or $1/N$ expansions of the
generating functional for one-particle irreducible ($1PI$) Green's functions
beyond leading order can be secular (unbounded) and break down at late 
\mbox{times \cite{LObeyond}}. It is known that the
calculation of transport coefficients \cite{Jeon}  
involves already an infinite series of diagrams not included at any 
finite order in the $1/N$ expansion of the $1PI$ effective action. 
In contrast, the relevant diagrams can be efficiently captured in 
finite-order \mbox{loop \cite{Calzetta2000}} 
or $1/N$ \mbox{expansion \cite{Berges:2001fi}} at NLO and NNLO 
of the $2PI$ effective action. The three-loop order has been also
frequently used as a starting point for kinetic descriptions  
and comprises the Boltzmann equation 
\cite{KadanoffBaym,Danielewicz,Calzetta:1988cq,Mrowczynski:1990bu,Blaizot:2001nr,Ivanov:2000tj}.

\section{Thermal fixed point}

\begin{figure}[t]
\begin{center}
\vspace*{-0.5cm}
\epsfig{file=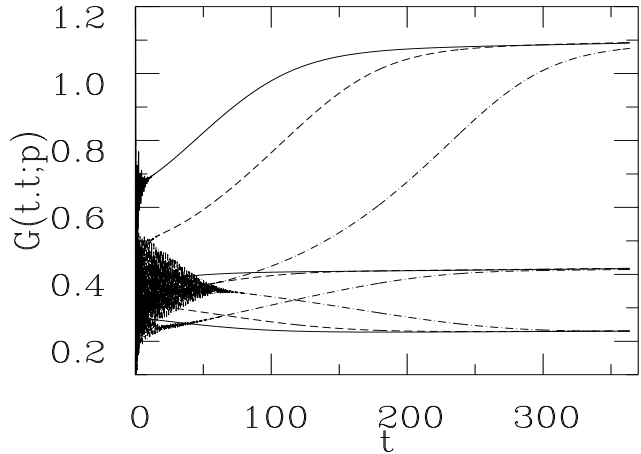,width=7.cm,height=11.5cm,angle=90}
\end{center}
\vspace*{-0.8cm}
\caption{Examples for the time dependence of the equal-time 
propagator $G(t,t;\bp)$ with Fourier modes $|\bp|=0,3,5$ 
from a first $2PI$ three-loop solution$^{4}$
(all in initial-time mass units). The evolution is
shown for three very different 
initial conditions with the same average energy density. After an
effective damping of rapid oscillations at early times the modes
are still not close to equilibrium. It is followed by a smooth
``drifting'' of modes and a subsequent late-time approach to
thermal equilibrium. The late-time
behavior of the correlator modes is insensitive to the
initial conditions and their value uniquely determined 
by the conserved energy density. This qualitative behavior has been 
verified beyond the loop/weak coupling expansion 
using the ($2PI$) $1/N$ expansion at next-to-leading order$^6$.}
\label{fixedpoint}
\end{figure}
We consider a scalar quantum field theory with action,
\beq
\label{classical}
S = \int {d}^{d+1}x\, \Big( \frac{1}{2} 
\partial_{x^0} \varphi_a
\partial_{x^0} \varphi_a
-\frac{1}{2} \partial_{\bx} \varphi_a
\partial_{\bx} \varphi_a
- \frac{1}{2} m^2 \varphi_a \varphi_a
- \frac{\lambda}{4! N} \left(\varphi_a\varphi_a\right)^2 \Big)
\eeq
where $\varphi_a(x)$, $x\equiv (x^0,\bx)$, is a real, 
scalar field with $a=1,\ldots,N$ components. For $N=4$
this corresponds to the linear sigma model for the light
scalar and pseudoscalar pion degrees of freedom $(\sigma,\vec{\pi})$
respecting global chiral $O(4)$ symmetry, or to the Higgs sector
of the electroweak standard model in absence of gauge field interactions. 
To save computational time
the numerical results will be presented for $1\!+1\!$ dimensions.
We stress that in the low dimensional theory many realistic
questions cannot be addressed, in particular since there is no 
spontaneous symmetry breaking. The 
dynamics may help to understand the symmetric regime relevant for 
sufficiently high energy densities.

Before describing the nonequilibrium dynamics of this
model in more detail below we give here an overview. 
As pointed out in Ref.\ \cite{Berges:2000ur}, for very different nonequilibrium
initial conditions a universal asymptotic late-time
behavior of quantum fields can be observed.
This behavior is demonstrated
in Fig.\ \ref{fixedpoint} for the equal-time propagator $G(t,t;\bp)$
in the three-loop $2PI$ approximation 
as a function of time $t$ for three Fourier modes, each starting from 
three very different initial
conditions with the same average energy density.
For the solid line the initial conditions are close
to a mean field thermal solution, the initial mode 
distribution for the dashed and the dashed-dotted lines deviate 
more and more from thermal equilibrium\cite{Berges:2000ur}. It is striking 
to observe that propagator modes 
with very different initial values but with the same momentum $\bp$ 
approach the same large-time value\footnote{We emphasize that 
thermalization or independence of the initial 
conditions cannot be observed in a strict sense because of 
time-reversal invariance (note that thermal equilibrium is invariant 
under time translation and bears no information about
initial conditions). Our results demonstrate that thermal
equilibrium can be approached closely with time, 
indistinguishable for practical matters. It cannot 
reach it on a fundamental level without some kind of coarse graining
which is a matter of principle and not a question of  
approximation.}. 
The late-time result approaches
the equilibrium correlation modes of the corresponding thermal
field theory \cite{Berges:2000ur}.

The observed real-time dynamics of correlator modes bears some
similarities with a Wilsonian renormalization group flow of couplings
towards an attractive fixed point. For the latter the large-distance
or low-energy behavior near the fixed point is insensitive to details 
at short distances or high energies.
For the real-time evolution we observe that the large-time behavior
is insensitive to details at early times. The role of the fixed point
is played here by a time-translation invariant solution for
the correlators obeying the standard periodicity (KMS) condition
for thermal equilibrium. 
Time translation invariant solutions 
play an important role for the late-time dynamics of nonequilibrium 
field theory. This fact is known as well for the leading order (LO) 
large-$N$ or mean field type approximations \cite{LOapp,LOapp2,LOapp3} 
which neglect scattering: 
At LO the correlator modes approach time-translation invariant solutions 
at asymptotically large times \cite{Test,LOasym,Berges:2001fi}. 
Those leading order ``fixed points''\cite{Wetterich:1997rp}
are distinct from the thermal one. In particular, the
LO fixed points depend explicitly on the initial 
particle number distribution, 
the simple reason beeing that the LO approximation
exhibits an additional conserved 
quantity (particle number) which is not present in the full theory. 
The LO fixed points become unstable once scattering is taken into 
account and the system approaches the thermal fixed  
(cf.\ the detailed LO/NLO comparison in Ref.\ \cite{Berges:2001fi}). 
We emphasize that
the described qualitative aspects are the same if we take into account 
scattering in the ($2PI$) three-loop or $1/N$ expansion 
at NLO, though quantitative aspects are different.          
We now go through the different characteristic time regimes
in more detail using the $1/N$ expansion of the $2PI$ effective
action at NLO, which provides for sufficiently large $N$
a small nonperturbative expansion parameter.

\section{Controlled nonperturbative approach}

All correlation functions 
of the quantum theory can be obtained from the effective 
action $\Gamma[\phi,G]$, here the $2PI$ generating functional
for Green's functions, which is 
parametrized by the field $G_{ab}(x,y)$ representing the
expectation value of the time ordered composite
$T \varphi_a(x) \varphi_b(y)$ 
and the macroscopic field $\phi_a(x)$ given by the expectation 
value of $\varphi_a(x)$.\cite{Cornwall:1974vz} 
We will concentrate in the following on the
symmetric regime where it is sufficient to consider 
$\Gamma[\phi=0,G] \equiv \Gamma[G]$.
The $2PI$ generating functional for Green's 
functions can be parametrized as \cite{Cornwall:1974vz}
\be
\Gamma[G] = \frac{i}{2} \Tr\ln G^{-1} 
          + \frac{i}{2} \Tr\, G_0^{-1} G
          + \Gamma_2[G] + {\rm const},  
\label{2PIaction}
\ee 
where $G_0^{-1}=i(\square+m^2)$ denotes the free inverse propagator.   
Writing  
$
\Gamma_2[G]= \Gamma_2^{\rm LO}[G] + \Gamma_2^{\rm NLO}[G] +\ldots
$
the LO and NLO contributions are given by
\cite{Berges:2001fi,Mihaila:2001sr}
\bea
&&\Gamma_2^{\rm LO}[G] = - \frac{\lambda}{4! N} 
  \int_{\C} {\rmd}^{d+1}x\, G_{aa}(x,x) G_{bb}(x,x), 
\label{LOcont} 
\\
&&\Gamma_2^{\rm NLO}[G] =  \frac{i}{2} \int_{\C} {\rmd}^{d+1}x\, 
\ln [\, {\bf B}(G)\, ] (x,x).
\label{NLOcont} 
\eea
Here $\C$ denotes the Schwinger-Keldysh contour along the real
time axis \cite{Schwinger:1961qe}
and 
\beq
{\bf B}(x,y;G) = \delta_{\C}^{d+1}(x-y)
+ i \frac{\lambda}{6 N}\, G_{ab}(x,y)G_{ab}(x,y).
\label{Vertex}
\eeq
The nonlocal four-point vertex at NLO is given by  
$\frac{\lambda}{6N}\, {\bf B}^{-1}$. \cite{Berges:2001fi}
In absence of external sources the evolution equation for $G$ is
determined by \cite{Cornwall:1974vz}
\beq
\frac{\delta \Gamma[G]}{\delta G_{ab}(x,y)} = 0.
\label{stationary}
\eeq
In the following we solve Eq.~(\ref{stationary}),
using Eqs.~(\ref{2PIaction})--(\ref{Vertex})
without further approximations numerically in $1\!+1\!$ dimensions.
For a detailed description of the approach,
the numerical implementation and
results see Ref.\ \cite{Berges:2001fi}.
We present the results using the decomposition identity 
\beq
G(x,y) = F(x,y)- (i/2) \rho(x,y)\, {\rm sign}_{\C}(x^0-y^0)
\eeq
where $F$ is the symmetric or statistical two-point function and $\rho$
denotes the spectral function \cite{Aarts:2001qa}. In thermal 
equilibrium $F$ and $\rho$ would be related by the fluctuation-dissipation
theorem, however, far from equilibrium both two-point functions
are linearly independent.

\section{Early-time exponential damping}

\begin{figure}[t]
\begin{center}
\epsfig{file=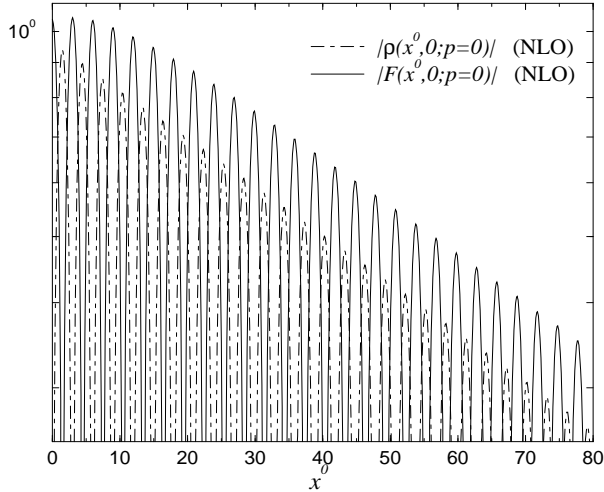,width=7.9cm,height=6.7cm,angle=0}
\end{center}
\vspace*{-0.6cm}
\caption{
Logarithmic plot$^6$ of the spectral and statistical two-point 
modes, $|\rho(x^0,0;p=0)|$
and $|F(x^0,0;p=0)|$ (in units of $M_{\rm I}$). 
The correlation modes oscillations quickly approach an exponentially
damped behavior with a corresponding characteristic time scale. 
\vspace*{-0.1cm}
} 
\label{FigLogFrho}
\end{figure} 
We study the time evolution for two classes of initial condition 
scenarios away from equilibrium. The first one corresponds to 
a ``quench'' where initially at high temperature one considers 
the relaxation processes following an instant ``cooling'' described 
by a sudden drop in the effective mass. 
The second scenario is characterized by initially Gaussian distributed
modes in a narrow momentum range around $\pm p_{\rm ts}$, which is
reminiscent of colliding wave packets moving with opposite and equal 
momentum \cite{Tsunami}. A similar nonthermal and radially
symmetric distribution of highly populated modes may also be encountered in a 
``color glass condensate'' at saturated gluon density with typical
momentum scale $p_{\rm ts}$ \cite{McLerran:1994ni}.

We start with a quench from an initial high temperature  
particle number distribution $n_0(p)=1/(\exp[\sqrt{p^2+M_0^2}/T_0]-1)$ 
with $T_0=2 M_{\rm I}$, $M_0^2= 2 M_{\rm I}^2$ in
units of the renormalized mass $M_{\rm I}$ at initial time $t=0$,
and a weak effective coupling $\lambda/6N = 0.083 \, M_{\rm I}^2$ 
for $N=10$. From Fig.\ \ref{FigLogFrho} one observes that
the statistical and spectral two-point functions, $F$ and $\rho$, 
oscillate with initial frequency $\epsilon_0/2\pi = 0.17 M_{\rm I}$.
The shown correlation functions quickly approach an exponentially damped
behavior with characteristic rate $\gamma^{\rm (damp)}_0 =0.016 M_{\rm I}$.
After the damping time scale 
$\tau^{\rm (damp)} \sim 1/\gamma_0^{\rm (damp)}$
correlations with initial times are effectively
suppressed and asymptotically $F$ and $\rho(t,0;p) \to 0^+$. 

\begin{figure}[t]
\begin{center}
\epsfig{file=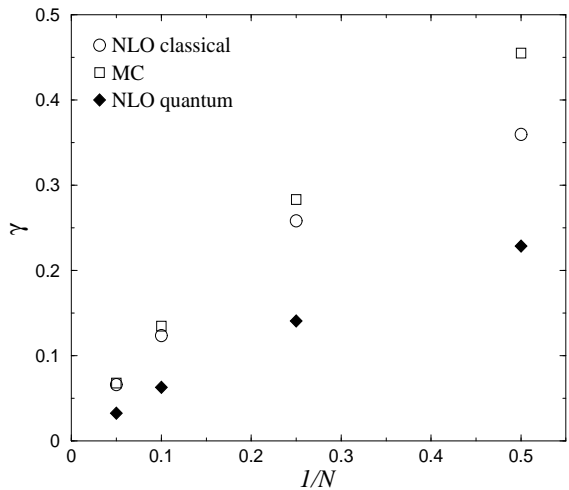,width=7.5cm,height=6.7cm,angle=0}
\end{center}
\vspace*{-0.6cm}
\caption{Nonequilibrium damping rates extracted from $F(t,0;p=0)$ 
as a function of $1/N$ for strong initial coupling 
$\lambda/M_{\rm I}^2=30$. The quantum results are shown 
with full symbols. The damping rates are reduced compared to the classical
field theory limit$^7$ \mbox{(open symbols)}. 
%The damping time
%scales $\sim N$ for large $N$. 
For the classical theory 
round symbols represent results from the
next-to-leading order $1/N$ approximation and square ones
denote the exact MC results.
One observes a rapid convergence of the $1/N$
expansion at NLO to the exact result already for moderate
values of $N$.
\vspace*{-0.3cm}  
} 
\label{fig2}
\end{figure}
In Fig.\ \ref{fig2} we show the damping rate as a function
of $1/N$, this time for a strong initial coupling 
$\lambda/M_{\rm I}^2=30$. The initial particle number 
$n_0(p)$ is chosen to  
represent a Gaussian distribution peaked around 
$p_{\rm ts}=2.5 M_{\rm I}$ with a thermal background of
temperature $T_0=4 M_{\rm I}$ \cite{Berges:2001fi,AB2}. The full symbols
show the NLO results for the quantum theory. One observes
that the damping time $\sim 1/\gamma^{\rm damp}$ scales
proportional to $N$ for sufficiently large $N$. It becomes
apparent that LO ($N \to \infty$) or mean field type 
approximations fail to describe damping correctly and
can be valid only for times $t \ll \tau^{\rm (damp)}$.
To demonstrate 
the power of the ($2PI$) $1/N$ expansion 
we repeat the NLO calculation for the corresponding classical 
statistical field theory where comparison
with exact (MC) results \cite{AB2}, including all orders of $1/N$,
is possible. The systematic convergence of the NLO and the Monte Carlo
result is apparent from Fig.~\ref{fig2}.

\vspace*{-0.1cm}  
\section{Non-exponential/power law drifting at intermediate times}

\vspace*{-0.1cm}  
After the characteristic early-time scale 
$\tau^{\rm (damp)}\sim 1/\gamma^{\rm damp}$
the system is typically still far away from 
equilibrium for a relatively long time. 
For a large variety of initial conditions we find a 
subsequent parametrically slow, 
non-exponential ``drifting'' of modes, 
as exemplified in Fig.\ \ref{fixedpoint}.
\begin{figure}[t]
\begin{center}
\epsfig{file=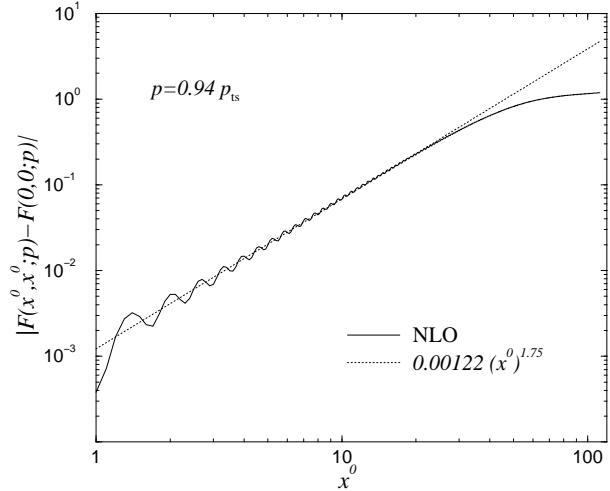,width=7.9cm,height=6.7cm,angle=0}
\end{center}
\vspace*{-0.6cm}
\caption{Double logarithmic plot$^6$ for the equal-time mode
$F(x^0,x^0;p)$ for an initial particle number distribution
peaked around $p = p_{\rm ts}$.
The dotted straight line corresponds to a power law behavior
$\sim (x^0\, M_{\rm INIT})^{1.75}$. 
At later times the behavior changes to an exponential 
approach to thermal equilibrium with 
rate $\gamma^{\rm therm}$ $\ll$ $\gamma^{\rm damp}$.
\vspace*{-0.2cm}
} 
\label{Figptsu}
\end{figure}
To become more quantitative we show in Fig.\ \ref{Figptsu}
the statistical two-point function on a double logarithmic
plot for a Gaussian initial particle number distribution
similar to the one above and with small effective coupling 
$\lambda/6N = 0.1 M_{\rm I}^2$ for \mbox{$N=10$. \cite{Berges:2001fi}} 
Time-averaged over the oscillation period
the evolution of the equal-time mode is well 
approximated by a power law behavior for times $t \lesssim 30/M_{\rm INIT}$. 
The observed scaling behavior is reminiscent of a cascade in fully developed 
turbulence. However, note that the initial condition is
homogeneous in space and all the energy is initially concentrated
in a small momentum range. The energy has then to be 
distributed from the densely populated modes at high momentum 
in particular to the low momentum modes in order to reach thermal 
equilibrium. The presence of power law behavior at intermediate times 
can be observed for a variety of initial conditions. For higher 
$n$-point functions one can also observe approximate scaling, 
however, the detailed
time evolution before the late-time thermalization can be rather
complex and depends on the initial conditions. 
See Ref.\ \cite{Berges:2001fi} for more details. 

\vspace*{-0.2cm}
\section{Late-time exponential thermalization}

\vspace*{-0.1cm}
We emphasize that irrespective of the details of the  
initial conditions we \mbox{find \cite{Berges:2001fi}} quite different
rates for damping and for the late-time exponential
approach to thermal equilibrium. For the example of 
Fig.\ \ref{Figptsu} the characteristic 
time scales $\gamma^{\rm (damp)}/\gamma^{(\rm therm)} \sim 
{\cal O}(10)$.\footnote{For 
the employed spatially homogeneous initial conditions we do not
observe a power-law ``tail'' for the late-time evolution. This
behavior is insensitive to changes of large volumes.}
\begin{figure}[t]
\begin{center}
\epsfig{file=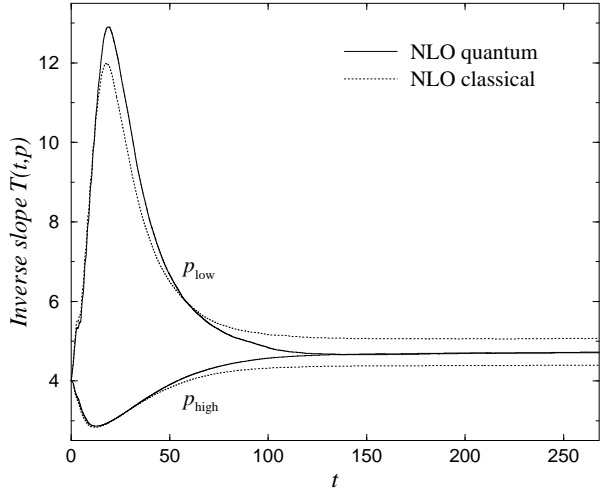,width=7.9cm,height=6.7cm,angle=0}
\end{center}
\vspace*{-0.6cm}
\caption{Time dependence of the inverse slope $T(t,p)$, as defined in the
text$^{7,6}$. When quantum thermal equilibrium  with a Bose-Einstein
distributed particle number $n(t,\epsilon_p)$ is approached, all 
modes have to get equal $T(t,p)=T_{\rm eq}$. 
In contrast, for classical thermal equilibrium the inverse
slope is momentum dependent with $T(p_{\rm low}) > T(p_{\rm high})$. 
}
\label{fig4}
\end{figure}
Though the exponential damping at early times
is crucial for an effective loss of details of the initial conditions
--- a prerequisite for the approach to equilibrium --- 
it does not determine the time scale for thermalization.
It is possible to relate the damping rate 
to the ``width'' of the Wigner transformed spectral function $\rho$
in the same way as discussed in \mbox{Ref.\ \cite{Aarts:2001qa}}, however, 
the thermalization rate cannot. 
This fact may be particularly pronounced
in $1\!+1\!$ dimensions where on-shell two-to-two 
scattering is constrained by the energy conservation relation such
that it does not change the particle numbers for the involved momentum
modes. Processes with nontrivial momentum exchange are necessary 
for thermalization and are taken into account at NLO by
off-shell effects. Correspondingly the calculation of the three-particle 
peak from off-shell decay of the nonequilibrium spectral
function in Ref.\ \cite{Aarts:2001qa} clearly shows a nonvanishing 
contribution. On-shell particle number changing processes
appear at NNLO and may change
quantitative aspects. The on-shell corrections beyond NLO are known
to be small in the statistical field theory limit where exact results
including all orders in $1/N$ show very good agreement with the NLO 
approximation already for moderate values of $N$.\ \cite{AB2} 

The nontrivial fact that quantum thermal equilibrium is approached
becomes more pronounced if one compares with classical \cite{Test} 
thermalization.  The classical field approximation
is expected to become a reliable description for the quantum theory if
the number of field quanta in each field mode is sufficiently high. 
Accordingly, we observe that increasing the initial particle number 
density leads to a convergence of quantum and classical time 
\mbox{evolution \cite{AB2}}. However, since classical and quantum 
thermal equilibrium are distinct the respective evolutions
have to deviate at sufficiently late times, irrespective of the
initial particle number density.
Differences in the particle number distribution can be conveniently
discussed using the inverse slope parameter
$T(t,p) \equiv - n(t,\epsilon_p) \mbox{$[n(t,\epsilon_p)+1]$} 
(dn/d\epsilon)^{-1}$
for a given time-evolving particle number distribution $n(t,\epsilon_p)$
and dispersion relation $\epsilon_p(t)$.\cite{Berges:2001fi}
Following Ref.\ \cite{Aarts:2001qa} we define the effective
particle number as     
$
n(t,\epsilon_p)+\frac{1}{2}
\equiv [F(t,t';p)\, \partial_{t}\partial_{t'} 
F(t,t';p) ]^{1/2}|_{t=t'}
$ 
and mode energy by
$
\epsilon_p(t) \equiv [\partial_{t}\partial_{t'} 
F(t,t';p)/F(t,t';p)]^{1/2}|_{t=t'}
$,
which coincide with the usual free-field definition for $\lambda \to 0$. 
For a Bose-Einstein distributed particle number the parameter $T(t,p)$
corresponds to the (momentum independent) temperature $T(t,p)=T_{\rm eq}$.  
In the classical limit the inverse slope $T(t,p)$ as
defined above remains momentum dependent.
In Fig.~\ref{fig4} we plot the function
$T(t,p)$ for $p_{\rm low}\simeq 0$ and $p_{\rm high} \simeq 2 p_{\rm ts}$. 
Initially one observes  
a very different behavior of $T(t,p)$ for the low
and high momentum modes, indicating that the system is far from
equilibrium. Note that classical and quantum evolution agree very well for
sufficiently high initial particle number density \cite{AB2}.
However, at later times the quantum evolution approaches
quantum thermal equilibrium with a constant inverse slope 
$T=4.7 M_{\rm I}$.\cite{Berges:2001fi} 
In contrast, in the classical limit the slope parameter remains momentum
dependent and the system relaxes towards classical thermal 
equilibrium \cite{AB2}.

\section{Conclusions}

The $2PI$ effective action provides 
a powerful framework to find practicable approximations for 
nonequilibrium dynamics. Combined
with a ($2PI$) $1/N$ expansion beyond leading order one obtains
a controlled nonperturbative description of far-from-equilibrium dynamics
at early times as well as late-time thermalization.
This includes the important capability of the $1/N$ expansion to 
describe nontrivial scaling properties near second order phase 
transitions at NLO and beyond. Most pressing is the extension
to $3\!+1\!$ dimensions along the lines of Ref.\ \cite{Berges:2001fi}.
This allows one to quantitatively address in a quantum 
field theory the formation of disoriented chiral
condensates or fluctuations near critical points in the context 
of heavy-ion collisions, or of (p)reheating
at the end of inflation in the early universe.

\section*{Acknowledgments}
\noindent
I thank my collaborators G.\ Aarts and J.\ Cox, and
B.\ M{\"u}ller for helpful discussions.
Many thanks to the organizers of this very interesting workshop and
for the invitation to give this talk.


\section*{References}
\begin{thebibliography}{99}

%\cite{Bodeker:2001pa}
\bibitem{Bodeker:2001pa}
For a review see 
D.\ B\"odeker,
%``Non-equilibrium field theory,''
Nucl.\ Phys.\ Proc.\ Suppl.\ {\bf 94} (2001) 61.
%[hep-lat/0011077].      

%\cite{Cornwall:1974vz}
\bibitem{Cornwall:1974vz}
J.~M.~Cornwall, R.~Jackiw, E.~Tomboulis,
%``Effective Action For Composite Operators,''
Phys.\ Rev.\ {\bf D10} (1974) 2428; see also
J.M.\ Luttinger and J.C.\ Ward, Phys.\ Rev.\ {\bf 118} (1960) 1417;
\mbox{G.\ Baym}, Phys.\ Rev.\ {\bf 127} (1962) 1391.
%%CITATION = PHRVA,D10,2428;%%          

\bibitem{Calzetta:1988cq}
E.~Calzetta, B.~L.~Hu,
%``Nonequilibrium Quantum Fields: Closed Time Path Effective Action,
%Wigner Function And Boltzmann Equation,''
Phys.\ Rev.\ {\bf D37} (1988) 2878;
K.~Chou, Z.~Su, B.~Hao and L.~Yu,
%``Equilibrium And Nonequilibrium Formalisms Made Unified,''
Phys.\ Rept.\ {\bf 118} (1985)~1.


%\cite{Berges:2000ur}
\bibitem{Berges:2000ur}
J.~Berges, J.~Cox,
%``Thermalization of quantum fields from time-reversal invariant evolution
%equations,'' 
Phys.\ Lett.\ {\bf B} to appear [hep-ph/0006160].
%%CITATION = HEP-PH 0006160;%%  

%\cite{Aarts:2001qa}
\bibitem{Aarts:2001qa}
G.~Aarts, J.~Berges,
%``Nonequilibrium time evolution of the spectral function in quantum field
%theory,''
Phys.\ Rev.\ {\bf D64} (2001) 0850XX [hep-ph/0103049].
%%CITATION = HEP-PH 0103049;%%   

%\cite{Berges:2001fi}
\bibitem{Berges:2001fi}
J.~Berges,
%``Controlled nonperturbative dynamics of quantum fields out of
%equilibrium,'' 
Nucl.\ Phys.\ {\bf A} to appear
[hep-ph/0105311].
%%CITATION = HEP-PH 0105311;%%

\bibitem{AB2} G.\ Aarts, J.\ Berges, hep-ph/0107129.

%\cite{Mihaila:2001sr}
\bibitem{Mihaila:2001sr}
B.~Mihaila, F.~Cooper, J.~F.~Dawson,
%``Resumming the large-N approximation for time evolving quantum
%systems,''
Phys.\ Rev.\ {\bf D63} (2001) 096003 (2001).
%%CITATION = HEP-PH 0006254;%%

%\cite{Blagoev:2001ze}
\bibitem{Blagoev:2001ze}
K.~Blagoev, F.~Cooper, J.~Dawson, B.~Mihaila,
%``Schwinger-Dyson approach to non-equilibrium classical field theory,''
hep-ph/0106195.
%%CITATION = HEP-PH 0106195;%%

\bibitem{LOinh} For 
improved results using inhomogeneous mean fields see 
G.~Aarts, J.~Smit,
%``Particle production and effective thermalization in inhomogeneous mean
%field theory,''
Phys.\ Rev.\ {\bf D61} (2000) 025002;
M.~Sall\'e, J.~Smit, J.~C.~Vink,
%``Thermalization in a Hartree ensemble approximation to quantum field
%dynamics,''hep-ph/0012346.
Phys.\ Rev.\ {\bf D64} (2001) 025016.
L.~M.~Bettencourt, K.~Pao, J.~G.~Sanderson,
%``Dynamical behavior of spatially inhomogeneous relativistic  
%lambda phi**4 quantum field theory in the Hartree approximation,''
hep-ph/0104210.

\bibitem{Andreas} For an early example see A.\ Ringwald, Phys.\ Rev.\ 
{\bf D36} (1987) 2598.

\bibitem{LObeyond} 
L.M.A.\ Bettencourt, C.\ Wetterich, Phys.\ Lett.\ {\bf B430}
(1998) 140; 

\bibitem{LObeyondOsc} For studies in quantum mechanics see
B.~Mihaila, J.~F.~Dawson, F.~Cooper,
%``Order 1/N corrections to the time-dependent Hartree 
%approximation for a  system of N+1 oscillators,''
Phys.\ Rev.\ {\bf D56} (1997) 5400;
L.~M.~Bettencourt, C.~Wetterich,
%``Time evolution of correlation functions for classical 
%and quantum  anharmonic oscillators,''
hep-ph/9805360; 
B.\ Mihaila, T.\ Athan, F.\ Cooper, J.\ Dawson, S.\ Habib, 
%``Exact and approximate dynamics of the quantum mechanical O(N) model,''
Phys.\ Rev.\  {\bf D62} (2000) 125015; 
A.~V.~Ryzhov, L.~G.~Yaffe,
%``Large N quantum time evolution beyond leading order,''
Phys.\ Rev.\  {\bf D62} (2000) 125003.

\bibitem{Jeon} \mbox{S.\ Jeon}, Phys.\ Rev.\ {\bf D52} (1995) 3591; 
S.\ Jeon and L.\ Yaffe, Phys.\ Rev.\ {\bf D53} (1996) 5799. 

\bibitem{Calzetta2000} 
E.A.\ Calzetta, B.L.\ Hu, S.A.\ Ramsey, Phys.\ Rev.\ {\bf D61}
(2000) 125013. 

%\cite{KadanoffBaym}
\bibitem{KadanoffBaym}
L.P.~Kadanoff, G.~Baym, 
``Quantum Statistical Mechanics'',
Benjamin, New York (1962).

\bibitem{Danielewicz}
P.\ Danielewicz, Ann.\ Phys.\ {\bf 152} (1984) 239;
{\bf 197} (1990) 154;

%\cite{Mrowczynski:1990bu}
\bibitem{Mrowczynski:1990bu}
S.~Mrowczynski, P.~Danielewicz,
%``Green Function Approach To Transport Theory Of Scalar Fields,''
Nucl.\ Phys.\ {\bf B342} (1990) 345;
S.~Mrowczynski and U.~Heinz,
%``Towards a relativistic transport theory of nuclear matter,''
Annals Phys.\ {\bf 229} (1994) 1.
%%CITATION = APNYA,229,1;%%

%\cite{Blaizot:2001nr}
\bibitem{Blaizot:2001nr}
J.~Blaizot, E.~Iancu,
%``The quark-gluon plasma: Collective dynamics and hard thermal loops,''
hep-ph/0101103.

%\cite{Ivanov:2000tj}
\bibitem{Ivanov:2000tj}
Y.~B.~Ivanov, J.~Knoll, D.~N.~Voskresensky,
%``Resonance Transport and Kinetic Entropy,''
Nucl.\ Phys.\ {\bf A672} (2000) 313;
%[nucl-th/9905028];
%%CITATION = NUCL-TH 9905028;%%
%\cite{Leupold:2000ga}
%\bibitem{Leupold:2000ga}
S.~Leupold,
%``Towards a test particle description of transport processes for states
%with continuous mass spectra,''
Nucl.\ Phys.\ {\bf A672} (2000) 475.

\bibitem{Test}
G.\ Aarts, G.F.\ Bonini, C.\ Wetterich,
%``Exact and truncated dynamics in nonequilibrium field theory,''
Phys.\ Rev.\ D {\bf 63} (2001) 025012.

\bibitem{LOapp} 
F.\ Cooper, S.\ Habib, Y.\ Kluger, E.\ Mottola, J.P.\ Paz,
P.R.\ Anderson, Phys.\ Rev.\ {\bf D50} (1994) 2848;
F.\ Cooper, S.\ Habib, Y.\ Kluger, E.\ Mottola, Phys.\ Rev.\ {\bf D55} 
(1997) 6471.

\bibitem{LOapp2}
D.\ Boyanovsky, H.J.\ de Vega, R.\ Holman, D.S.\ Lee, A.\ Singh,
Phys.\ Rev.\ {\bf D51} (1995) 4419; D.\ Boyanovsky, H.J.\ de Vega, 
R.\ Holman, J.F.J.\ Salgado, Phys.\ Rev.\ {\bf D54} (1996) 7570;
D.~Boyanovsky, H.~J.~de Vega, R.~Holman and J.~Salgado,
Phys.\ Rev.\ {\bf D59} (1999) 125009; 

\bibitem{LOapp3}
J.~Baacke, K.~Heitmann and C.~P\"atzold,
Phys.\ Rev.\ {\bf D55} (1997) 2320, Phys.\ Rev.\ {\bf D57} (1998) 6406;
J.\ Baacke, S.\ Michalski, hep-ph/0109137.

%asymptotic behavior:
\bibitem{LOasym}
D.\ Boyanovsky, C.\ Destri, H.J.\ de Vega, 
R.\ Holman, J.\ Salgado, Phys.\ Rev.\ {\bf D57} (1998) 7388;

%\cite{Wetterich:1997rp}
\bibitem{Wetterich:1997rp}
C.~Wetterich,
%``Nonequilibrium time evolution in quantum field theory,''
Phys.\ Rev.\  {\bf E56} (1997) 2687 ; Phys.\ Rev.\ Lett.\  {\bf 78}
(1997) 3598.

%\cite{Schwinger:1961qe}
\bibitem{Schwinger:1961qe}
J.~Schwinger,
%``Brownian Motion Of A Quantum Oscillator,''
J.\ Math.\ Phys.\ {\bf 2} (1961) 407;
%%CITATION = JMAPA,2,407;%%
%\cite{Keldysh:1964ud}
%\bibitem{Keldysh:1964ud}
L.~V.~Keldysh,
%``Diagram Technique For Nonequilibrium Processes,''
Zh.\ Eksp.\ Teor.\ Fiz.\ {\bf 47} (1964) 1515 
[Sov.\ Phys.\ JETP {\bf 20} (1965) 1018].
%%CITATION = ZETFA,47,1515;%%

\bibitem{Tsunami}
R.D.\ Pisarski, hep-ph/9710370;
%``Nonabelian Debye screening, tsunami waves, and worldline fermions,''
D.\ Boyanovsky, H.J.\ de Vega, R.\ Holman, S.\ Prem Kumar, R.D.\ Pisarski,
%``Non-equilibrium evolution of a *tsunami*: Dynamical symmetry breaking,''
Phys.\ Rev.\ {\bf D57} (1998) 3653. 

\bibitem{McLerran:1994ni}
L.~McLerran, R.~Venugopalan,
%``Computing quark and gluon distribution functions for very large nuclei,''
Phys.\ Rev.\  {\bf D49} (1994) 2233; (1994) 3352; for a recent review see 
L.~McLerran,
%``The color glass condensate and small x physics: 4 lectures,''
hep-ph/0104285 and references therein; R.D.\ Pisarski, private communication.


\end{thebibliography}
\end{document}